\documentclass[10pt, conference]{IEEEtran}

\usepackage{amsmath,amssymb,amsfonts}
\usepackage{graphicx}
\usepackage{textcomp}

\usepackage{amsthm}
\usepackage{xcolor}
\usepackage[figuresright]{rotating}

\usepackage{graphicx}
\graphicspath{{/}{fig/}}

\usepackage{array}
\usepackage{textcomp}
\usepackage{xcolor}
\usepackage{multirow}
\usepackage{booktabs}

\usepackage{mathtools}
\usepackage{breqn}

\usepackage{caption}
\usepackage{mathtools}
\usepackage{float}
\usepackage{hyperref}
\usepackage{flushend}

\usepackage{pgfplots}

\newlength\figureheight
\newlength\figurewidth
\title{\LARGE \bf
    UWB-Based Localization for Multi-UAV Systems and Collaborative Heterogeneous Multi-Robot Systems: a Survey
}

\author{
    \IEEEauthorblockN{
        Wang Shule\textsuperscript{1,2},
        Carmen Mart\'{i}nez Almansa\textsuperscript{2},
        Jorge Pe\~{n}a Queralta\textsuperscript{2},
        Zhuo Zou\textsuperscript{1},
        Tomi Westerlund\textsuperscript{2}
    }\\
    \IEEEauthorblockA{
        \textsuperscript{1} School of Information Science and Technology, Fudan University, Shanghai, China \\
        \textsuperscript{2} \href{https://tiers.utu.fi}{Turku Intelligent Embedded and Robotic Systems Lab, University of Turku, Turku, Finland} \\
        Emails: \textsuperscript{1}zhuo@fudan.edu.cn \textsuperscript{2}\{shwang, camart, jopequ, tovewe\}@utu.fi \\
    }
}

\begin{document}

\maketitle
\thispagestyle{empty}
\pagestyle{empty}

\begin{abstract}

    Ultra-wideband technology has emerged in recent years as a robust solution for localization in GNSS denied environments. In particular, its high accuracy when compared to other wireless localization solutions is enabling a wider range of collaborative and multi-robot application scenarios, being able to replace more complex and expensive motion-capture areas for use cases where accuracy in the order of tens of centimeters is sufficient. We present the first survey of UWB-based localization focused on multi-UAV systems and heterogeneous multi-robot systems. We have found that previous literature reviews do not consider in-depth the challenges in both aerial navigation and navigation with multiple robots, but also in terms of heterogeneous multi-robot systems. In particular, this is, to the best of our knowledge, the first survey to review recent advances in UWB-based (i) methods that enable ad-hoc and dynamic deployments; (ii) collaborative localization techniques; and (iii) cooperative sensing and cooperative maneuvers such as UAV docking on mobile platforms. Finally, we also review existing datasets and discuss the potential of this technology for both localization in GNSS-denied environments and collaboration in multi-robot systems.

\end{abstract}


\IEEEpeerreviewmaketitle

\section{Introduction}

Accurate localization is an essential aspect of the navigation of autonomous robots in GNSS-Denied environments. Existing approaches can be classified among those relying on on-board sensors only, such as visual or lidar odometry~\cite{qingqing2019odometry}, or fixed elements in the environment in known locations, such as visual markers~\cite{kong2013autonomous}, or wireless beacons~\cite{faragher2014analysis}. In the case of aerial robots, which have gained momentum over the past decade, accurate localization poses additional challenges due to the three-dimensional nature of their operational environment. Outdoors, unmanned aerial vehicles (UAVs) often rely on GNSS sensors for localization and autonomous navigation~\cite{stempfhuber2011precise}. However, in multiple application scenarios only GNSS-denied navigation is possible, from emergency and post-disaster situations~\cite{cui2015drones}, to navigation in industrial environments such as factories or warehouses~\cite{tiemann2017scalable}. Multiple challenges remain in the utilization of odometry methods based on onboard sensors for accurate and robust localization in long-term autonomy~\cite{qingqing2019odometry}, while the main drawback of most existing wireless localization solutions is their lower accuracy~\cite{faragher2014analysis}. One of the advantages of wireless methods is that they are less dependent on the environment, such as in low-visibility conditions or the presence or dust of smoke. More recently, ultra-wideband (UWB) localization systems have emerged as high-accuracy solutions, in the order of tens of centimeters or even centimeters~\cite{queralta2020uwb}, based on wireless radio modules. In UWB-based localization systems, the position of a UWB node on a mobile robot can be calculated based on ranging measurements to fixed UWB nodes located in known positions. UWB-based has the potential to replace more complex and expensive motion-capture arenas based on visual markers in application scenarios where a localization accuracy of the order of tens of centimeters is sufficient~\cite{furtado2019comparative}. An additional benefit of UWB-based systems in such scenarios is the larger area that can be covered, with recent commercial modules being able to communicate up to a distance of 60\,m in line-of-sight. Finally, UWB signals enable non-line-of-sight localization at similar accuracy levels~\cite{6071927}.

\begin{figure}
    \centering
    \includegraphics[width=0.3923\textwidth]{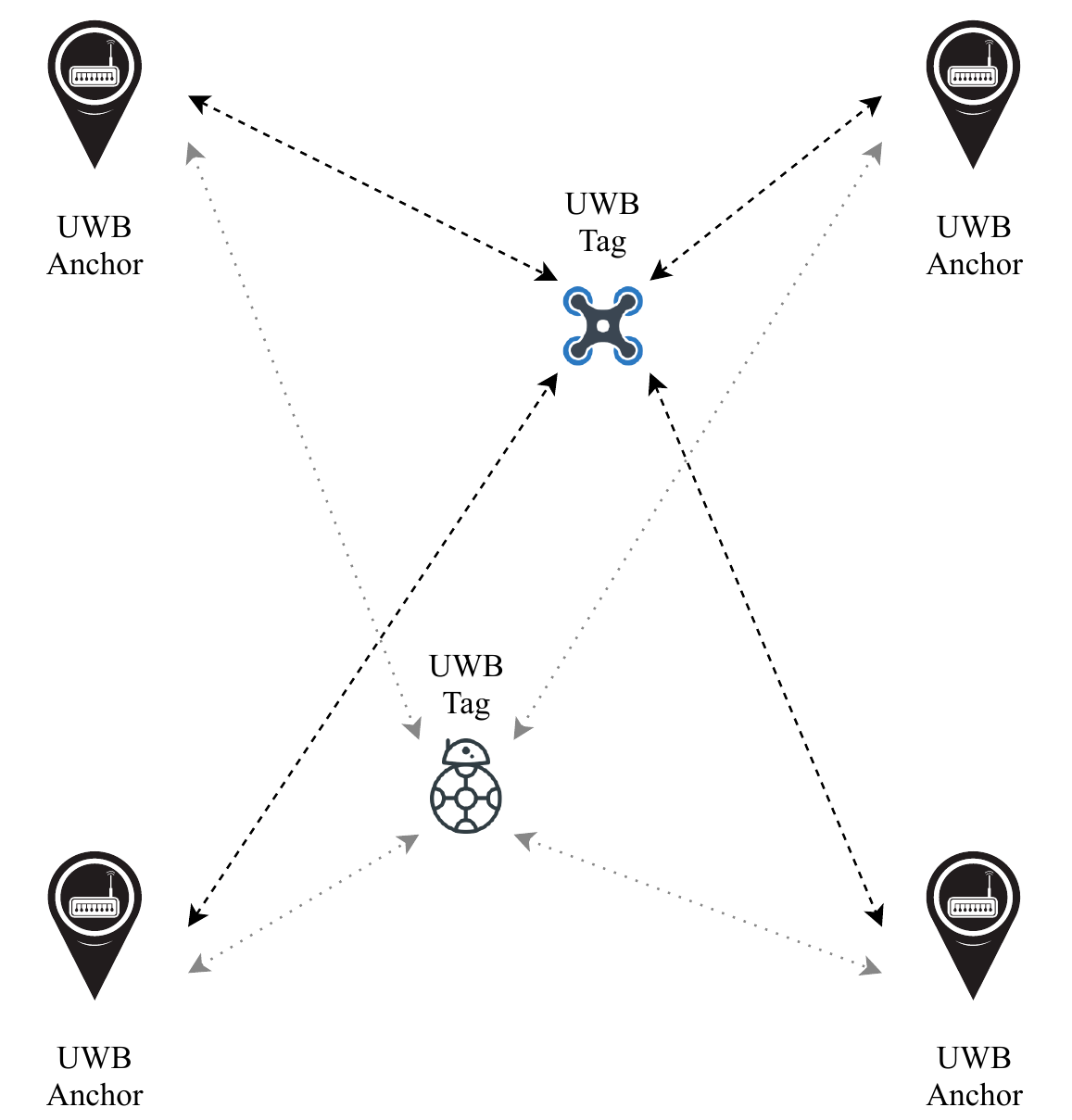}
    \caption{Multiple robots with UWB tags can be localized in real-time based on individual ranging measurements to the fixed system anchors.}
    \label{fig:concept}
\end{figure}

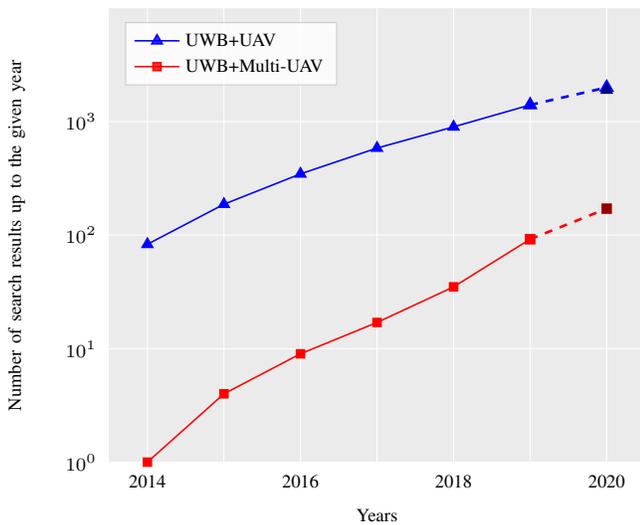
\begin{figure}
    \centering
    \setlength{\figureheight}{0.42\textwidth}
    \setlength{\figurewidth}{0.48\textwidth}
    \scriptsize{
\begin{tikzpicture}

\begin{axis}[
    axis lines=left,
    axis line style={draw=none},
    tick style={draw=none},
    height=\figureheight,
    width=\figurewidth,
    axis background/.style={fill=white!92!black},
    legend cell align={left},
    legend style={fill opacity=0.8, draw opacity=1, text opacity=1, at={(0.03,0.97)}, anchor=north west, draw=white!80!black},
    x grid style={white!72!white},
    xlabel={Years},
    xmajorgrids,
    xmin=0.5, xmax=7.5,
    xtick style={color=black},
    xtick={1,2,3,4,5,6,7},
    xticklabels={2014,,2016,,2018,,2020},
    y grid style={white!72!white},
    ylabel={Number of search results up to the given year},
    ymajorgrids,
    ymin=0, ymax=9900,
    ytick style={color=black},
    ymode=log,
]
\addplot [semithick, blue, mark=triangle*, mark size=2.3, mark options={solid}]
table {%
1 83
2 187
3 346
4 584
5 898
6 1401
};
\addlegendentry{UWB+UAV}
\addplot [line width=1pt, blue, dashed, mark=triangle*, mark size=2.3, mark options={solid}, forget plot]
table {%
6 1401
7 1997
};
\addplot [semithick, red, mark=square*, mark size=1.5, mark options={solid}]
table {%
1 1
2 4
3 9
4 17
5 35
6 92
};
\addlegendentry{UWB+Multi-UAV}
\addplot [line width=1pt, red, dashed, mark=square*, mark size=1.5, mark options={solid}, forget plot]
table {%
6 92
7 171
};
\addplot [semithick, blue!54.5098039215686!black, dashed, mark=triangle*, mark size=2.3, mark options={solid}, forget plot]
table {%
7 1897
};
\addplot [semithick, red!54.5098039215686!black, dashed, mark=square*, mark size=1.5, mark options={solid}, forget plot]
table {%
7 171
};
\end{axis}

\end{tikzpicture}}
    \caption{Number of publications in Google Scholar that mention UWB for UAV and Multi-UAV systems from 2014 to 2019 and predicted number for 2020.}
    \label{fig:pubs}
\end{figure}

Multiple works have applied UWB for localization in sensor networks within the Internet of Things (IoT) systems~\cite{chen2020uwb}, as well as ground robots or people in indoor environments~\cite{raza2019dataset}. In this paper, however, we focus on surveying works that have applied UWB for localization in UAVs, and in particular in multi-UAV systems and heterogeneous multi-robot systems. In these cases, the localization problem poses additional challenges, and multiple research questions remain open. Moreover, this survey aims at filling the gap in review papers in this field, which has seen a fast growth over the past few years. The number of scientific publications reporting the utilization of UWB-based localization systems for UAVs has increased exponentially over the last 5 years, as illustrated in Fig.~\ref{fig:pubs}. In the figure, we report the total number of results from Google Scholar, where the search has been carried out for any paper including all of these three keywords in their text: "UWB" (or ultra-wideband), "UAV" and "localization". When searching over papers including both "UWB" and "UAV" in the title, a total of 28 results appear from the period 2014--2020, out of which 10 ($\sim$36\%) are from 2019 onward, and 17 ($\sim$61\%) from 2018 onward. In summary, we see a clear trend in the research of UWB-based localization for UAVs, while no survey has covered the different use cases and approaches so far, to the best of our knowledge. The only related review paper to date was carried out by F. Zafari \textit{et al.}, where the authors described different localization methods for GNSS-denied environments and UWB systems played a minor role~\cite{zafari2019survey}. Nonetheless, in their paper the authors already point out the potential of UWB for high-accuracy localization and the rising adoption of UWB-based systems across different domains.

UWB-based localization systems have been utilized in spectacles and commercial UAVs. For instance, Verity Studios~\cite{cirque2016zurich} utilizes an undisclosed UWB-based approach for the localization of multi-UAV systems. In 2019, drones relying on UWB for localization were deployed for a human-drone dance~\cite{moscow2019uwb}. In terms of commercial UAVs, Bitcraze's has developed a UWB add-on for the Crazyflie drone~\cite{giernacki2017crazyflie}. The add-on, called Loco positioning deck, utilizes Decawave's DWM1000 UWB transceiver for indoor positioning. Decawave is one of the main companies in the area, providing multiple commercial products and ready-to-use bundles for UWB localization based on their real-time localization system (DRTLS). Other companies providing both UWB radio transceivers and higher-end solutions with ready-to-use systems are Pozyx, Sewio and OpenRTLS~\cite{contigiani2016implementation}. The rest of the paper is structured as follows. Section 2 introduces the basic concepts behind UWB-based localization systems. Section 3 then focuses on their applicability for UAV navigation in indoor environments and, in particular, multi-UAV systems. Section 4 discusses UWB for heterogeneous multi-robot systems, while Section 5 lists the existing related datasets. Section 7 concludes the work.

\section{Ultra-Wideband Localization Systems}

In this section we introduce the different methods utilized for measuring ranging information between pairs or groups of UWB transceivers, and estimating the position of mobile UWB tags based on the known position of fixed UWB anchors. First, we discuss the options for estimating the distance between the emitter and receiver of a UWB signal. Second, we describe the different existing localization systems, including cooperative localization methods.
%
%
\subsection{Measurement modalities}

The distance between a UWB transceiver emitting a signal and the receiver node can be estimated from the time of flight of the signal and then utilizing the known speed of transmission of electromagnetic waves in air. 

\begin{figure}
    \centering
    \includegraphics[width=0.42\textwidth]{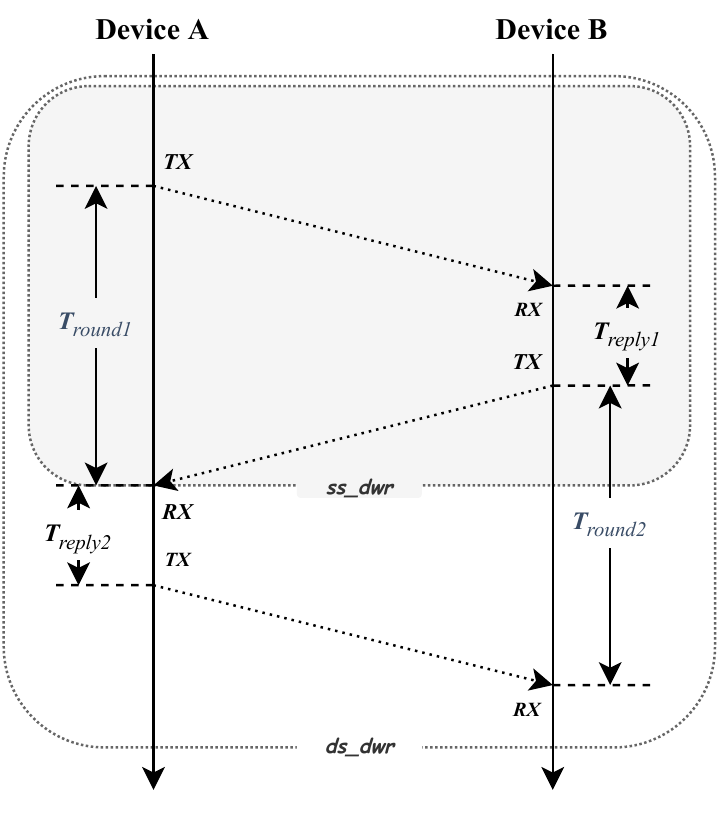}
    \caption{Multiple robots with UWB tags can be localized in real-time based on individual ranging measurements to the fixed system anchors.}
    \label{fig:twr}
\end{figure}

\noindent\textit{(i) Time of Flight (ToF)}, or Time of Arrival (ToA), calculates the propagation time of the wireless signal from the emitter node to the receiving node by recording the sending and receiving timestamps of the ranging message, multiplying the speed of light and then obtaining the distance between the devices~\cite{ni2019uwb,yang2017kfl}. Depending on whether the devices are synchronized or not, ToF can be one-way ranging or two-way ranging. In one-way ranging, the ranging message only propagates in one direction and both devices must maintain accurate clock synchronization. In two-way ranging, clock synchronization is not needed and therefore it is a more widely utilized approach, with lower overall complexity. The TWR (Two-Way Ranging) method requires two-way communication between devices, which again depending on the information available about the antenna delays and packet processing latency, can be divided into two methods: Single-Sided Two-Way Ranging (SS-TWR) and Double-Sided Two-Way Ranging (DS-TWR)~\cite{wang2019design}. These two methods are illustrated in Fig.~\ref{fig:twr}. In the SS-TWR algorithm, device A initiates a ranging request message, while device B responds to the ranging and returns a message processing delay $T_{reply}$. After receiving the response message, device A calculates the round-trip delay $T_{round}$ of the message, and then the flight time between nodes A and B can be calculated from~(\ref{eq:ss-twr}). 

\begin{equation}
    T_{prop}=0.5*{\left ( T_{round}- T_{reply}\right )}
    \label{eq:ss-twr}
\end{equation}

In the DS-TWR algorithm, both devices A and B will initiate a ranging request, which is equivalent to the SS-TWR initiated by both devices. Then, the flight time between nodes A and B can be calculated according to (\ref{eq:ds-twr}). Because no previous calibration is needed, in general it yields than SS-TWR.

\begin{equation}
    T_{prop}=\frac{T_{round1}*T_{round2}-T_{reply1}*T_{reply2}}{T_{round1}+T_{round2}+T_{reply1}+T_{reply2}}
    \label{eq:ds-twr}
\end{equation}

\noindent\textit{(ii) Time Difference of Arrival (TDoA)}, also known as hyperbolic positioning, measures the difference in the propagation time between the UWB signal from a transmitting UWB tag and two or more receiving UWB anchors, to then obtain the relative difference in distance between the tag and each of the anchors~\cite{martalo2019uwb, zhao2019uwb}. In general, it can improve the localization accuracy over ToA. However, TDoA algorithm does not directly use the difference between the signal emission and arrival time; instead, it uses the time difference between the signals received by multiple UWB anchors to calculate the position of a moving UWB tag~\cite{cheng2019uwb}. While it does not need to add a special time signal for clock synchronization between tags and anchors, anchors must be either interconnected or synchronized~\cite{wang2019design}.

\subsection{Localization System Setup}

Most existing papers utilizing UWB-based systems for localization in multi-robot systems rely on a set of fixed anchors with known positions and constant height~\cite{8250979, 4380919, 8768568}. In these settings, a least squares estimator (LSE) and a Kalman filter are commonly used to track the position of the tags. Multiple works have focused on increasing the localization accuracy, often by fusing the UWB ranging data with other sensors. In this direction, Lee \textit{et al.} presented an IR-UWB system to enhance the positioning performance with ToF measurements~\cite{8250979}. Krishnan \textit{et al.} proposed a similar IR-UWB approach for TDoA-based localization~\cite{4380919}. For NLOS transmissions, Hyun \textit{et al.} relied on a ray-tracing algorithm given the footprint of the area where the nodes where moving~\cite{8768568}. Other authors have proposed IMU/UWB fusion~\cite{6249434}, vision-UWB fusion~\cite{7838839}, or lidar-UWB fusion~\cite{8990648}.


\subsubsection{Mobile Setting for Ad-Hoc Deployments}

An early approach to a mobile UWB localization system was proposed by K. C. Cheok \textit{et al.} to track mobile robots~\cite{5671780}. The authors utilized a mobile reference platform where they placed four UWB modules at points with known coordinates. A fifth UWB transceiver was placed on a separate mobile robot. The robot's position is then obtained by trilateration. The novelty of their solution is that it includes a system that calibrates the anchors automatically. The algorithm used by that system calculates the coordinates of each anchor from the distances measured between them, taking the first anchor as the origin of the coordinate system. This automatic calibration makes the solution's setup faster and easier compared to those where the anchors' coordinates have to be defined manually. The limitations of trilateration, such as noise, were overcome in their experiments by using additional sensors and an Extended Kalman Filter (EKF) to fuse all the measurements gathered. The main limitation was the accuracy of the UWB ranging itself, as the technology was not so mature at the moment.

More recently, M. Hamer \textit{et al.} presented a multiple-robot localization system also based on UWB auto-calibrated modules~\cite{8344407}. Their deployment consists of an extensible network of eight UWB anchors that allows several robots to localize themselves within the anchors' area simultaneously. The robots use time difference of arrival (TDoA) to localize themselves, this requires clock synchronization in the anchor network. To authors also designed a Kalman filter to track the anchor positions over time. Another recent multi-robot system for localization based on UWB sensors was presented by S. Güler \textit{et al.}~\cite{8814678}. Their system consists of two robots and four UWB modules, three of them on the anchor robot and one on the target robot. The UWB devices on the anchor robot are used to estimate the relative position of the target robot. Their solution allows real-time measurements without communication with a central station or between robots. The main drawbacks of this configuration are the limited positioning of the anchor sensors, restricted by the size of the robot, which decreases the system performance, and the noisy measurements. To remove the latter they use a Monte-Carlo localization (MCL) algorithm based on a particle filter.

Finally, in~\cite{almansa2020autocalibration}, an autocalibration method for a mobile anchor system is introduced. The authors simulate a mobile system with four anchors and three tags, even though it can be extended to an arbitrary number of UWB nodes. The simulations were done based on experimental one-to-one measurements with Decawave's DWM1001 UWB module. The authors showed that, compared to the built-in autocalibration process from Decawave, their proposed system is able to significantly reduce the amount of calibration time while increasen the accuracy of the anchor positions.

\subsubsection{Collaborative/Distributed Localization}

Rather than relying on anchors to localize mobile tags individually, a different approach is to calculate the distances between the subsets of nodes in the networks and utilize collaborative localization algorithms. In this direction, B. Denis \textit{et al.} proposed a collaborative solution for synchronization and localization in UWB ad hoc networks~\cite{1618619}. Several anchors and many mobile nodes compose their experimental network. For synchronization, the TWR-TOA technique is used as well as a diffusion algorithm. The localization algorithm is a distributed maximum log-likelihood algorithm which reduces the NLOS ranging errors effect on accuracy. However, they conclude that the solution's sensitivity has to be further studied. Similarly, A. Prorok \textit{et al.} introduced a probabilistic UWB error model to increase multi-robot localization performance~\cite{6071927}. Their experiments also demonstrate the improvement achieved by using a collaborative approach in the precision of multiple-robot localization systems. For this, they include an extended particle filter algorithm which allows collaboration between robots. However, their error model is still to be tested in real-time. Finally, A. Subramanian \textit{et al.} presented a distributed localization method based on signal strength measurements of UWB modules~\cite{1599728}. In their approach, initially only the anchor node has a known location. It then moves along a \textit{mobility path} broadcasting its position at each point to enable other nodes, as many as possible, to receive one or more of them and thus, estimate their own position and later, broadcast it along their own path. Their results show that the proposed solution is scalable in both area size and number of nodes. In order to avoid problems derived from the noise received along with the signal a Kalman filter is utilized.
\section{UWB in Multi-UAV Systems}

The use of unmanned aerial vehicles (UAVs) has been rapidly expanding across multiple aspects of our lives in recent years~\cite{franke2015civilian}. Compared to unmanned ground vehicles (UGVs), UAVs present additional challenges in terms of localization and navigation. For instance, while a UGV can easily maintain its position, this can become a challenging problem in UAVs if accurate localization is not possible. Most autonomous UAVs to date rely on GNSS sensors for global positioning~\cite{schultz2012method, kim2016cubature}. Nonetheless, other solutions must be utilized indoors or in other GNSS-denied environments~\cite{queralta2020uwb}. A widely used localization system with high accuracy are motion capture arenas relying on high-speed cameras and reflective markers~\cite{campoy2009computer, ramon2016visual, tang2018vision}. In visual positioning and navigation systems, the amount of image processing is significant and the average computer cannot complete the calculation~\cite{lu2018survey, balamurugan2016survey}, requiring expensive hardware to achieve real-time performance. Moreover, the deployment is relatively fixed with limited area coverage. Therefore, researchers looking for more flexible solutions, even if not with the same levels of accuracy, have been moving to new localization systems for UAVs and Multi-UAV systems based on ultra-wideband (UWB) wireless transceivers. Among these, T. M. Nguyen~\textit{et al.} presented a Multi-UAV positioning system based on Two-Way Time-Of-Flight Ultra-Wideband (UWB) technique that stands out~\cite{nguyen2016ultra}. They used the Extended Kalman Filter (EKF) estimate and developed a new method based on Non-linear Regression (NLR) for simultaneous localization for four UAVs.







\subsection{Formation Control}


One of the first applications of UWB in multi-UAV systems is formation control algorithms. Formation control refers to the design of multi-robot control systems for spatial coordination~\cite{queralta2019progressive}. In many cases, these rely only on distance measurements~\cite{oh2015formationsurvey}, but more often relative positioning between the robots is required~\cite{mccord2019progressive}. Some algorithms do not need accurate localization, and collision avoidance is often an art of the control scheme~\cite{queralta2019indexfree}.

UWB has already been utilized in formation control algorithms for multi-UAV systems. K. {Guo}~\textit{et al.}~\cite{8680745} and S.Q. {Cao}~\textit{et al.}~\cite{Cao2018/03} both proposed a distributed formation control scheme. In their work, a UWB ranging and communication network was used for relative localization (RL) estimation in two dimensions. From such a network, the distances and relative speed between the UAVs can be estimated.In~\cite{8680745}, the authors achieved formation flights with three UAVs. Finally, in~\cite{Cao2018/03} the authors realized a leader-follower formation flight with two UAVs. In this latter work, the authors also performed an outdoor positioning comparison between UWB and GPS, concluding that UWB can fully meet the requirements of relative positioning for formation control algorithms at a similar level than GNSS sensors.


\subsection{Application Scenarios}

The applications of UAVs are multiple and increasing by the day, with use cases ranging from industrial applications~\cite{macoir2019uwb,cunha2018ultra}, to search and rescue operations~\cite{sequeira2017search,garcia2019autonomous}, or entertainment shows~\cite{moscow2019uwb, cirque2016zurich}. In this section, we describe some of the most recent applications of UWB and UAVs across different domains.


\subsubsection*{Industrial Applications} 

Industrial drones account for a large proportion in the UAV market, with a wide range of applications including warehouse management~\cite{macoir2019uwb}, coal mine inspection~\cite{cunha2018ultra}, environmental mapping~\cite{billdal2018sensor}, metric reconstructions~\cite{masiero2017low}, logistics express~\cite{chang2018optimal}, forest fire prevention~\cite{moulianitis2018evaluation} and many other industry scenarios. 

For warehouse applications, N. Macoir~\textit{et al.} designed a multi-technology UWB localization MAC protocol that can be applied for UAV-based inventory management~\cite{macoir2019uwb}. The authors designed a protocol to optimize the energy consumption of UWB anchor nodes achieving a positioning accuracy is of 5\,cm. In a subsequent paper, N. Macoir~\textit{et al.} also validated the designed UWB system in two different scenarios: (i) automatic drone navigation for inventory management and (ii) positioning runners on an indoor track without infrastructure. These results demonstrate the viability and effectiveness of UWB-based localization systems for real-world industrial applications.

In the mapping field, A. Masiero~\textit{et al.} used a low-cost UWB technique for metric reconstruction of buildings~\cite{masiero2017low}. They fixed 8 UWB anchors at different heights on the building, and then fixed a UWB tag on UAV, utilizing UWB ranging for a UAV positioning system to determine the location of the UAV with high accuracy. This enabled a more efficient combination and fusion of images taken by the UAV camera while navigating the building, rendering the 3D reconstruction of the building. The authors compared the resulting model with Terrestrial Laser Scanner (TLS) data, validating the utilization of UWB for accurate localization.

In the field of mining, F. Cunha~\textit{et al.} presented and analyzed UWB radar technology as a robust sensing solution to inspect the underground coal mines environment~\cite{cunha2018ultra}. While not directly related to UWB ranging, the experiments show that the sensor can maintain high accuracy under environmentally harsh conditions. Another recent work relying on UWB radars was presented by Billdal~\textit{et al.}~\cite{billdal2018sensor}. The authors simulated a novel sensor for industrial UAV inspection based on UWB radar technology. A map of the environment was utilized to supplement the UWB radar sensor and expand its range of applications.


\subsubsection*{Shows/Entertainment}

A more recent application of UAVs has been their utilization in light shows both indoors and outdoors. At the 14th Moscow International Aerospace Expo (MAKS), a Chinese drone company: DAMODA, performed a "human-machine dance" with UWB-aided UAVs~\cite{moscow2019uwb}. The entire indoor UAV formation is based on UWB localization technology to achieve precise positioning and precise control. Another notable company utilizing undisclosed UWB technology for indoor shows is Verity Studios, a Swiss company that designs multi-UAV systems for indoor spectacles~\cite{cirque2016zurich}.





\subsubsection*{Search and Rescue}

UAVs are playing an increasingly important role in search and rescue operations owing to their advantages of navigation flexibility, adaptability to multiple scenarios and relative simplicity of their deployment. In this area, a patent of M. F. Sequeira~\textit{et al.} describes a UAV search and rescue system and method based on UWB transceivers~\cite{sequeira2017search}. The UWB transceivers are defined to be used for detecting and locating survivors in a search and rescue area and then transmit the location to a command center via a separate data link transceiver on the UAV. In general, search and rescue operations in post-disaster GNSS-denied environments would benefit from UWB for accurate localization even in poor visibility conditions where other sensors might suffer from downgraded performance~\cite{queralta2020uwb}. Regarding the use of UWB radars in the field, M. Garcia-Fernandez~\textit{et al.} proposed an autonomous airborne Synthetic Aperture Radar (SAR) imaging system of UAVs, which are mainly composed by a UWB radar and Real-Time Kinematic (RTK) positioning system~\cite{garcia2019autonomous}. With these technologies, drones can obtain 3D high-resolution radar images of the underground environment to detect buried hazards such as mines and improvised explosive devices.

\subsubsection*{Surveillance/Monitoring}

Even in environments where GNSS sensors can be utilized, UWB can aid in reducing the localization error and increasing the overall accuracy. For instance, Y.-C. Chen~\textit{et al.} designed a higher-precision UAV Traffic Management (UTM) system by combining UWB ranging measurement with GPS and LoRa WAN\cite{chen2020uwb}. They used Decawave DW1000 chips as UWB transceivers and distributed multiple anchor nodes around the buildings in the test area. A series of actual multi-DUT experiments showed that the system can reduce the error of outdoor relative position between devices under test (DUTs) from 6\,m to just 10\,cm. In urban scenarios, UWB can also aid in locating and monitoring UAVs. H.Deng~\textit{et al.} proposed a UAV target tracking system based on machine vision and UWB positioning technology~\cite{CN104777847A}. In this case, UWB transceivers are utilized in the UAV's positioning system to monitor the unmanned aircraft's trajectory in real-time and transmit the coordinate data to an FPGA processor for processing. 

Finally, UWB radars can also aid in environmental monitoring. A six-engine drone that used UWB radar to monitor hazardous objects was designed by M. Wattimena~\textit{et al.}~\cite{wattimena2018board}. The purpose of using UWB radar was to study underground layers, especially the structure of solid and liquid mediums. The authors tested the prototype to detect the boundary of two liquids in a tank, which was calculated with an average square error of just 1\,mm. 







\section{UWB in Heterogeneous Multi-Robot Systems}

Multi-modal sensor fusion in heterogeneous multi-robot systems can significantly increase the situational awareness of individual robots~\cite{queralta2020blockchain}. However, the data fusion mechanisms often need accurate relative localization between the robots. The high-accuracy short-range distance estimation that UWB enables can then be exploited in this direction. Deploying multiple UWB transceivers in each robot enables not only relative localization between each pair of robots in terms of position but also orientation. Nguyen \textit{et al.} demonstrated the viability of this idea by designing an extended Kalman filter for robust target-relative localization in a heterogeneous multi-robot system, where UWB was utilized for both ranging and communication~\cite{nguyen2018robust}. In utilizing UWB for relative positioning, one of the most relevant publications to date in this area is Nguyen \textit{et al.}'s work on the first autonomous docking of UAVs in a mobile UGV platform that relies on UWB localization for approaching the mobile docking station~\cite{nguyen2019integrated}. The final docking maneuvers, however, are based on onboard vision and known markers on the docking platform. While the size of the mobile platform was relatively large, 2~m long and 1.5~m wide, recent datasets with multiple anchor configurations show the viability of this idea maintaining relatively high accuracy~\cite{queralta2020uwb}. Other works have focused on utilizing UWB for specific maneuvers. In~\cite{fan2018fully}, a multi-robot collision avoidance scheme was developed and tested with UWB transceivers for global localization of the agents via deep reinforcement learning. Qiang \textit{et al.} developed a multi-robot localization platform built on top of the Robot Operating System (ROS)~\cite{qiang2017design}. Their platform, which can accommodate various types of robots, was then extended and applied to a formation control problem in~\cite{wei2019consensus}. 











\section{UWB Localization for Robotics Datasets}

The literature in UWB-based localization systems for UAVs and multi-UAV systems is extensive. Nonetheless, there is only a limited number of open datasets focused on the localization of either UAVs or multi-robot systems. The oldest datasets we have found provide only one-dimensional data from one-to-one distance measurements using two UWB nodes. Among these, Raza \textit{et al.} introduced a dataset for indoor localization where the focus was to study the advantages and limitations of both narrow-band and ultra-wideband time-of-flight distance measurement systems~\cite{raza2019dataset}. More recently, Li \textit{et al.} published a dataset from an autonomous UAV flight in a GNSS-denied environment that relied on UWB-aided navigation~\cite{li2018dronedataset}. In their paper, the authors proposed the fusion of UWB and inertial data with an Extended Kalman Filter (EKF) to achieve high-accuracy three-dimensional localization. Their dataset, however, contains data from a single flight with a single anchor setting. In contrast, Peña Queralta \textit{et al.} focus on the analysis of how different anchor configurations affect the accuracy of the localization~\cite{queralta2020uwb}. This latter dataset is the most extensive to date, to the best of our knowledge, with multiple anchor settings and several manual and autonomous flight. In this paper, the authors also discuss the possibilities of a mobile anchor setting and report an improvement in the calibration of anchor positions compared to the methods provided by Decawave for their DWM1001 UWB transceiver.

\section{Conclusion}

We have presented the first survey focused on UWB for multi-robot localization in GNSS-denied environments. We have focused on the applications in multi-UAV systems and in heterogeneous multi-robot systems as a basis for efficient collaborative behaviour. We have first discussed the different localization modalities, with fixed anchor settings being predominant to estimate the position of mobile tags, but also describing the possibilities of cooperative localization in distributed networks. Second, we have discussed the potential of UWB ranging as an inexpensive and robust solution for formation control in multi-UAV systems and localization of UAVs in general. In this area, UWB has already been applied in different industrial environments. Finally, we have reviewed recent papers in which UWB is the backbone for cooperative actions in heterogeneous multi-robot systems, such as UAV docking on a mobile platform. In summary, UWB has the potential to become a standard technology for relative positioning and ranging in multi-robot systems, having been applied to a wide variety of scenarios in recent years.

\section*{Acknowledgements}

This work was supported by the Academy of Finland's AutoSOS project with grant number 328755, by the NSFC grant No. 61876039, and the Shanghai Platform for Neuromorphic and AI Chip (NeuHeilium).

\bibliographystyle{unsrt}
\bibliography{main}

\end{document}